\documentclass[12pt]{article}
\usepackage{graphicx}
\usepackage[figuresright]{rotating}
\input{epsf}
\newcommand{\beq}{\begin{equation}}
\newcommand{\eeq}{\end{equation}}
\newcommand{\be}{\begin{eqnarray}}
\newcommand{\ee}{\end{eqnarray}}
\begin{document}
\rightline{RUB-TPII-13/99}
\rightline{hep-ph/9910464}
\vspace{.3cm}
\begin{center}
\begin{large}
{\bf Polarized antiquark flavor asymmetry in
Drell--Yan pair production} \\[1cm]
\end{large}
\vspace{1.0cm}
{\bf B.\ Dressler}$^{\rm a, 1}$,
{\bf K.\ Goeke}$^{\rm a, 2}$,
{\bf M.V.\ Polyakov}$^{\rm a, b, 3}$,
{\bf P.\ Schweitzer}$^{\rm a, 4}$,
\\[.1cm]
{\bf M.\ Strikman}$^{{\rm c, 5}, \ast}$, 
{\bf and C. Weiss}$^{\rm a, 6}$
\\[1.cm]
$^{a}${\em Institut f\"ur Theoretische Physik II,
Ruhr--Universit\"at Bochum, \\ D--44780 Bochum, Germany}
\\
$^{b}${\em Petersburg Nuclear Physics Institute, Gatchina, \\
St.Petersburg 188350, Russia}
\\
$^{c}${\em Pennsylvania State University,
University Park, PA 16802, U.S.A.}
\end{center}
\vspace{1.5cm}
\begin{abstract}
\noindent
We investigate the role of the flavor asymmetry of the nucleon's
polarized antiquark distributions in Drell--Yan lepton pair
production in polarized nucleon--nucleon collisions at
HERA (fixed--target) and RHIC energies. 
It is shown that the large polarized antiquark
flavor asymmetry predicted by model calculations in the
large--$N_c$ limit (chiral quark--soliton model) has a dramatic 
effect on the double spin asymmetries in high mass lepton pair 
production, as well as on the single spin
asymmetries in lepton pair production through $W^\pm$--bosons at
$M^2 = M_W^2$.
\end{abstract}
\vfill
\rule{5cm}{.2mm} \\
{\footnotesize $^{\rm 1}$ E-mail: birgitd@tp2.ruhr-uni-bochum.de} \\
{\footnotesize $^{\rm 2}$ E-mail: goeke@tp2.ruhr-uni-bochum.de} \\
{\footnotesize $^{\rm 3}$ E-mail: maximp@tp2.ruhr-uni-bochum.de} \\
{\footnotesize $^{\rm 4}$ E-mail: peterw@tp2.ruhr-uni-bochum.de} \\
{\footnotesize $^{\rm 5}$ E-mail: strikman@physics.psu.edu} \\
{\footnotesize $^{\rm 6}$ E-mail: weiss@tp2.ruhr-uni-bochum.de} \\
{\footnotesize $^{\ast}$ Alexander--von--Humboldt--Forschungspreistr\"ager}
\newpage
Drell--Yan (DY) lepton pair production in $pp$ or $pn$ collisions offers
one of the most direct ways to measure the antiquark distributions
in the nucleon. In particular, such experiments have recently
established a significant flavor asymmetry of
the unpolarized antiquark distributions,
$\bar u(x) - \bar d (x)$, see Ref.\cite{DYreview} for a review.
Since the amount of $\bar u(x) - \bar d (x)$ generated perturbatively
is very small, this provides unambiguous evidence for
an important role of nonperturbative effects in generating the sea
distributions. Other evidence is the large suppression of the
strange sea compared to the nonstrange one for $Q^2$ of the order of
a few ${\rm GeV}^2$.
It appears natural to invoke the chiral degrees of freedom for the
explanation of these effects.
Two competing mechanisms are currently being discussed.
One is due to scattering off pions generated via virtual
processes $N\rightarrow N +\pi$, $N\rightarrow \Delta +\pi$,
or $q\rightarrow q + \pi$ \cite{cloud}. With this mechanism one can in
principle generate a significant value of
$\bar u(x) - \bar d (x)$, although this requires one to 
consider virtual pion momenta up to $\sim 1\, {\rm GeV}$
and relies on fine-tuning of the parameters of the model; see 
Ref.\cite{KFS96} for a discussion.
Another mechanism emerges within the large--$N_c$ limit of QCD,
where the nucleon can be described as a chiral
soliton \cite{DPPPW96,PPGWW98,Dressler98}. This approach
allows for a fully quantitative description of the 
antiquark distributions essentially without free parameters,
and preserves all fundamental qualitative properties 
of the distribution functions, such as positivity, sum rules {\it etc}.
It describes well the data for $\bar u(x) - \bar d (x)$ \cite{Dressler98}.
\par
\begin{figure}[t]
\begin{center}
\includegraphics[width=10cm,height=10cm]{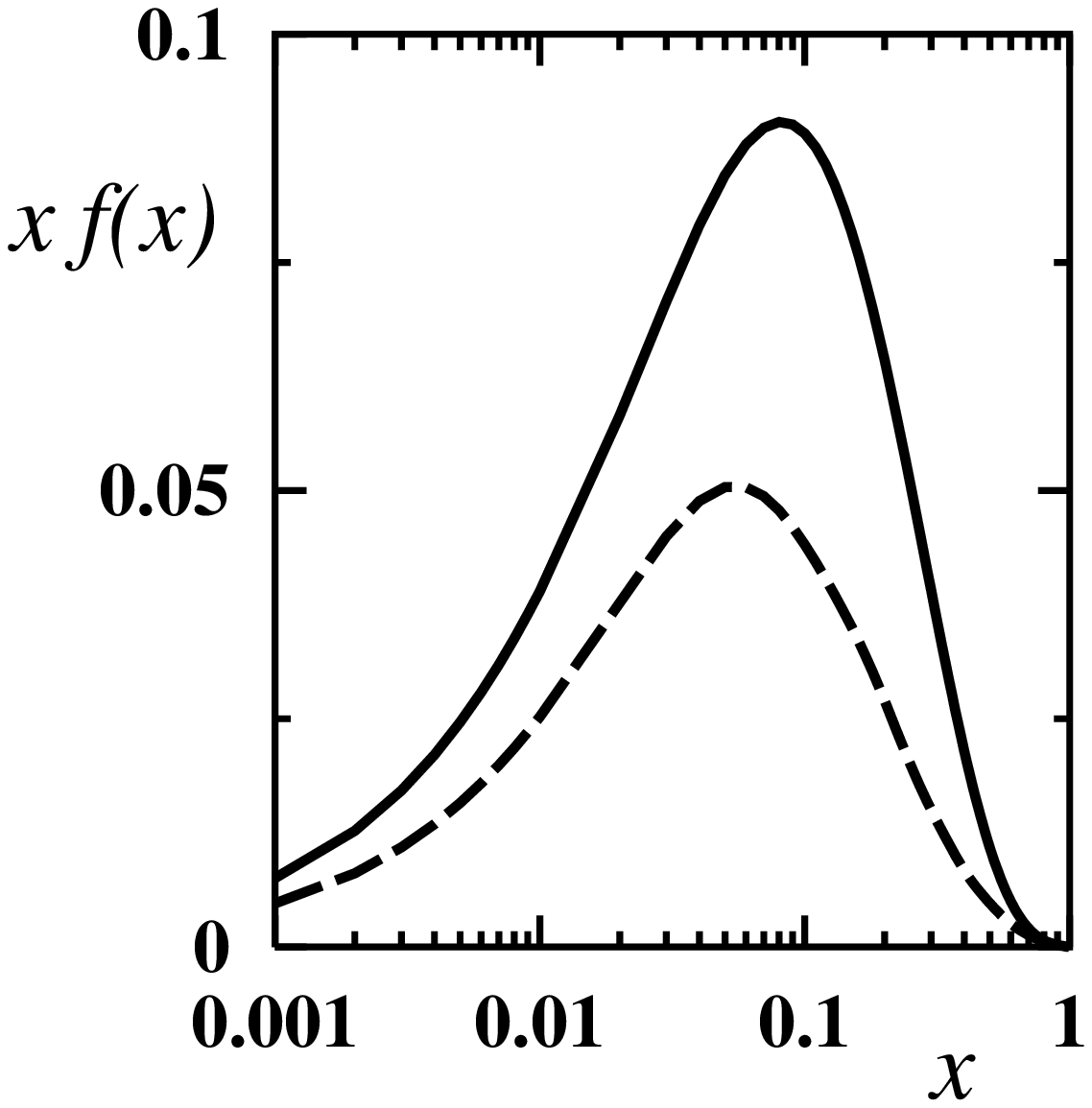}
\end{center}
\caption[]{The polarized and unpolarized antiquark flavor 
asymmetries obtained in model calculations in the large--$N_c$ 
limit (chiral quark--soliton model), evolved (LO) from the 
low normalization point of $\mu^2 = (600 \, {\rm MeV})^2$ 
to a scale of $\mu^2 = (5\, {\rm GeV})^2$.
\underline{Dashed line:} Unpolarized flavor asymmetry, 
$x [\bar d (x) - \bar u (x)]$, see Ref.{\rm \cite{PPGWW98}}.
\underline{Solid line:} Polarized flavor 
asymmetry, $x [\Delta \bar u (x) - \Delta \bar d (x)]
\equiv x \Delta_3 (x)$, 
see Refs.{\rm \cite{DPPPW96,Dressler99}}.}
\label{fig_unpol}
\end{figure}

It was pointed out in 
Ref.\cite{Dressler99} that a distinctive difference
of the two mechanisms is the degree of polarization of the
antiquark flavor asymmetry, $\Delta\bar u(x) - \Delta\bar d (x)$. 
In the pion cloud models polarization is
absent \cite{Zoller}. There have been some attempts
to generate polarization by including spin--$1$ resonances in this
picture \cite{Fries98}, which, however, presents
severe conceptual difficulties.\footnote{Pions play a special
role as the Goldstone bosons of spontaneously broken chiral symmetry.
In contrast, there is nothing special about exchanges of spin--1
resonances compared to, say, tensor, $b_1$, $h_1$,
$\rho_3$, $a_4$ {\it etc.}\ mesons. Moreover, Regge recurrences
are likely to lead to strong cancellations between contributions from 
different resonances. Also, the quark and gluon degrees
of freedom already partly account for the mesonic degrees of freedom, 
so one faces the problem of double counting. See 
Ref.\cite{Dressler99} for a critical discussion.}
In contrast to the pion cloud model
the large--$N_c$ approach predicts that
$\Delta\bar u(x) - \Delta\bar d (x)$ is much larger than the unpolarized
$\bar u(x) - \bar d (x)$; in fact, it is parametrically enhanced
by a factor of $N_c$. [The numerical results for the 
polarized \cite{DPPPW96,Dressler99} and 
unpolarized \cite{PPGWW98} antiquark flavor 
asymmetries obtained in this approach 
are shown in Fig.\ref{fig_unpol} at a scale of 
$\mu^2 = (5\, {\rm GeV})^2$.] Thus, measurements of
$\Delta\bar u(x) - \Delta\bar d (x)$ would provide a decisive test of
the different approaches to include the chiral degrees of freedom in
the nucleon.
\par
We have recently demonstrated that the current data on
hadron production in semi-inclusive deep--inelastic scattering (DIS)
are not sensitive to the value of  $\Delta\bar u(x) - \Delta\bar d (x)$
\cite{Dressler99}.
The purpose of this letter is to study if DY pair and
$W^\pm$ production in polarized $pp$ collisions, which will
be possible at RHIC, allow to distinguish between the two options.
Specifically, we investigate the role of the large
polarized antiquark flavor asymmetries obtained
in the large--$N_c$ model calculation of Ref.\cite{DPPPW96,Dressler99}
on spin asymmetries in longitudinally
polarized DY pair production.
\par
Predictions for the spin asymmetries in polarized DY pair
production (see {\it e.g.}\ Ref.\cite{Soffer98}) have so far been 
made on the basis of present experimental information about the 
polarized parton distributions in the nucleon,
which comes mostly from inclusive DIS \cite{GRSV96,GS96}.
However, DIS probes directly only the
sum of quark-- and antiquark distributions, while the separation
in quarks and antiquarks, as well as the gluon distribution,
have to be determined indirectly through scaling violations.
The flavor asymmetry of the polarized antiquark distribution is
practically not constrained by the
DIS data \cite{GRSV96,GS96}. On the other hand, the
polarized antiquark flavor asymmetry contributes to DY spin
asymmetries at leading order in QCD \cite{Kumano99}.
A quantitative understanding of these effects is a prerequisite
for any attempt to extract the polarized gluon
distribution from NLO analyses of the data \cite{Gehrmann97}.
\par
The cross section for DY pair production is a function of
the center--of--mass energy of the incoming hadrons,
$s = (p_1 + p_2)^2$, and the invariant mass of the produced
lepton pair, $M^2$, which is equal to the virtuality of the
exchanged gauge boson. At the partonic level this process 
is described by the annihilation of
a quark and an antiquark originating from the two hadrons,
carrying, respectively, longitudinal momenta $x_1 p_1$
and $x_2 p_2$, with $x_1 x_2 \; = \; Q^2 / s$.
One can parametrize the momentum fractions as
$x_1 \; = \; (Q^2 /s )^{1/2} e^y , \;\;
x_2 \; = \; (Q^2 /s )^{1/2} e^{-y}$, where $y$ is called rapidity.
In the case of DY pair production through a virtual
photon one is interested in the double spin asymmetry of the
cross section
\be
A^\gamma_{LL} &=& \frac{\sigma^\gamma_{++} - \sigma^\gamma_{+-}}
{\sigma^\gamma_{++} + \sigma^\gamma_{+-}} ,
\label{A_LL}
\ee
where the subscripts $+ , -$ denote the longitudinal polarization
of nucleons $1$ and $2$. In QCD in leading--log approximation this
ratio is given by \cite{Soffer98,Kamal98}
\be
A^\gamma_{LL}
(y; s, M^2 ) &=& \frac{ \sum_a e_a^2 \; \Delta q_a (x_1, M^2 )
\; \Delta q_{\bar a} (x_2 , M^2) }{ \sum_a e_a^2 \; q_a (x_1, M^2 )
\; q_{\bar a} (x_2 , M^2) } ,
\label{A_LL_parton}
\ee
where the sum runs over all species of light quarks and antiquarks
in the two nucleons, $a \; = \; \{ u, \bar u , d, \bar d , s , \bar s \}$; 
we neglect the small contributions due to heavy flavors.
The relevant scale here for the parton distribution functions
is the virtuality of the photon, $M^2$.
When the lepton pair is produced instead by exchange of a charged weak gauge
boson, $W^\pm$, due to the parity--violating nature of the weak
interaction the cross section exhibits already a single spin asymmetry,
\be
A^{W\pm}_L &=& \frac{\sigma^{W\pm}_{+} - \sigma^{W\pm}_{-}}
{\sigma^{W\pm}_{+} + \sigma^{W\pm}_{-}} ,
\label{A_L}
\ee
where now the subscripts $+ , -$ denote the longitudinal polarization of
nucleon $1$; the polarization of nucleon $2$
is averaged over. In QCD in leading--log approximation one 
has \cite{Soffer98,Kamal98}
\be
A^{W\pm}_{L} (y; s, M^2 )
&=&
\frac{  \Delta u(x_1, M^2 )\,\bar{d}(x_2, M^2 ) 
      - \Delta \bar{d}(x_1, M^2 )\,u(x_2, M^2 )}
     {  u(x_1, M^2 )\,\bar{d}(x_2, M^2 ) 
      +  \bar{d}(x_1 , M^2 )\,u(x_2 , M^2 )} ,
\label{A_L_parton}
\ee
for $W^-$ one should exchange $u \leftrightarrow d ,
\bar u \leftrightarrow \bar d$ everywhere here.
Eq.(\ref{A_L_parton}) includes only $u$-- and $d$--quarks, even 
for values of $M^2$ of the order of the $W$--boson mass. Contributions 
from $c$--$s$ transitions are negligible because 
of the comparative smallness of the product of $c$ and $s$ distributions, 
while contributions of type $u$--$s$ and $c$--$d$ are small because
of Cabbibo suppression; see Ref.\cite{Martin99} for a more detailed
discussion.
\par
Our aim is to study the effect of the large flavor asymmetry
of the polarized antiquark distributions, obtained in
the model calculations of Refs.\cite{DPPPW96,Dressler99}
based on the large--$N_c$ limit,
on the spin asymmetries $A^\gamma_{LL}$ and $A^{W\pm}_{L}$, 
Eqs.(\ref{A_LL_parton}) and (\ref{A_L_parton}).
In order to make maximum use of the direct experimental information
on the polarized parton distributions available from DIS we
proceed as follows. The individual polarized light quark and
antiquark distributions $\Delta u(x), \Delta \bar u(x),
\Delta d(x), \Delta \bar d(x), \Delta s(x)$, and $\Delta \bar s(x)$,
figuring in the numerators in Eqs.(\ref{A_LL_parton}) and
(\ref{A_L_parton}) can be expressed in terms of the
six combinations
\be
\Delta_u (x) &\equiv&
\Delta u (x) + \Delta \bar u(x),
\hspace{1cm} \mbox{(analogously for $\Delta_d, \Delta_s$)} ,
\label{Delta_u}
\\
\Delta_0 (x) &\equiv&
\Delta \bar u (x) + \Delta \bar d(x) + \Delta \bar s(x), \;\;\;
\label{Delta_0}
\\
\Delta_3 (x) &\equiv&
\Delta \bar u (x) - \Delta \bar d(x), \;\;\;
\label{Delta_3}
\\
\Delta_8 (x) &\equiv&
\Delta \bar u (x) + \Delta \bar d(x) - 2 \Delta \bar s(x) .
\label{Delta_8}
\ee
The combinations $\Delta_u (x) , \Delta_d (x) $ and $\Delta_s (x)$,
Eq.(\ref{Delta_u}), are measured directly in
inclusive polarized DIS, so we evaluate them using the
GRSV95 leading--order (LO) parametrization (``standard scenario''),
which was obtained by fits to inclusive DIS 
data \cite{GRSV96}.\footnote{Actually, in DIS with proton or nuclear targets
one is able to measure directly only two flavor combinations of these
three distributions; however, the third one can be inferred using
$SU(3)$ symmetry arguments.}
The flavor--singlet antiquark distribution, $\Delta_0 (x)$,
Eq.(\ref{Delta_0}), we also take from the GRSV95 parametrization;
this distribution is known only from the study of scaling violations
in inclusive DIS and depends to some extent on the assumptions made
about the polarized gluon distribution; however, the GRSV95
parametrization for $\Delta_0 (x)$ is in good agreement
with the result of model calculations in the large--$N_c$ limit
\cite{WK99}. For the polarized flavor asymmetries of the
antiquark distribution, $\Delta_3 (x)$ and $\Delta_8 (x)$,
Eqs.(\ref{Delta_3}) and (\ref{Delta_8}), which are not constrained
by DIS data, we use the results of the model calculation in the
large--$N_c$ limit of Refs.\cite{DPPPW96,Dressler99}, evolved 
in LO from the low normalization point of
$\mu^2 = (600\, {\rm MeV} )^2$ to the experimental scale, $M^2$.
The result for $\Delta_3 (x)$ is shown in Fig.\ref{fig_unpol}
at a scale of $(5\, {\rm GeV})^2$. The other non-singlet combination,
$\Delta_8 (x)$, is obtained from $\Delta_3 (x)$ at the low normalization
point by the $SU(3)$ relation 
$\Delta_8 (x) = [(3 F - D)/(F + D)] \Delta_3 (x)$, where we use
$F/D = 5/9$, see Ref.\cite{Dressler99} for details.
Note that $\Delta_3 (x)$ and $\Delta_8 (x)$ do not mix with the
other distributions under LO evolution.
The ``hybrid'' polarized quark and antiquark distributions thus
obtained, by construction, fit all the inclusive polarized DIS data
in LO, while at the same time incorporating the polarized antiquark
flavor asymmetry obtained in the model calculation in the large--$N_c$ 
limit. Finally, to evaluate the denominators in Eqs.(\ref{A_LL_parton})
and (\ref{A_L_parton}) we use the GRV94 parametrization
of the unpolarized parton distributions.
\par
\begin{figure}[t]
\begin{tabular}{rr}
\begin{tabular}{c}
\includegraphics[width=7.2cm, height=7.2cm]{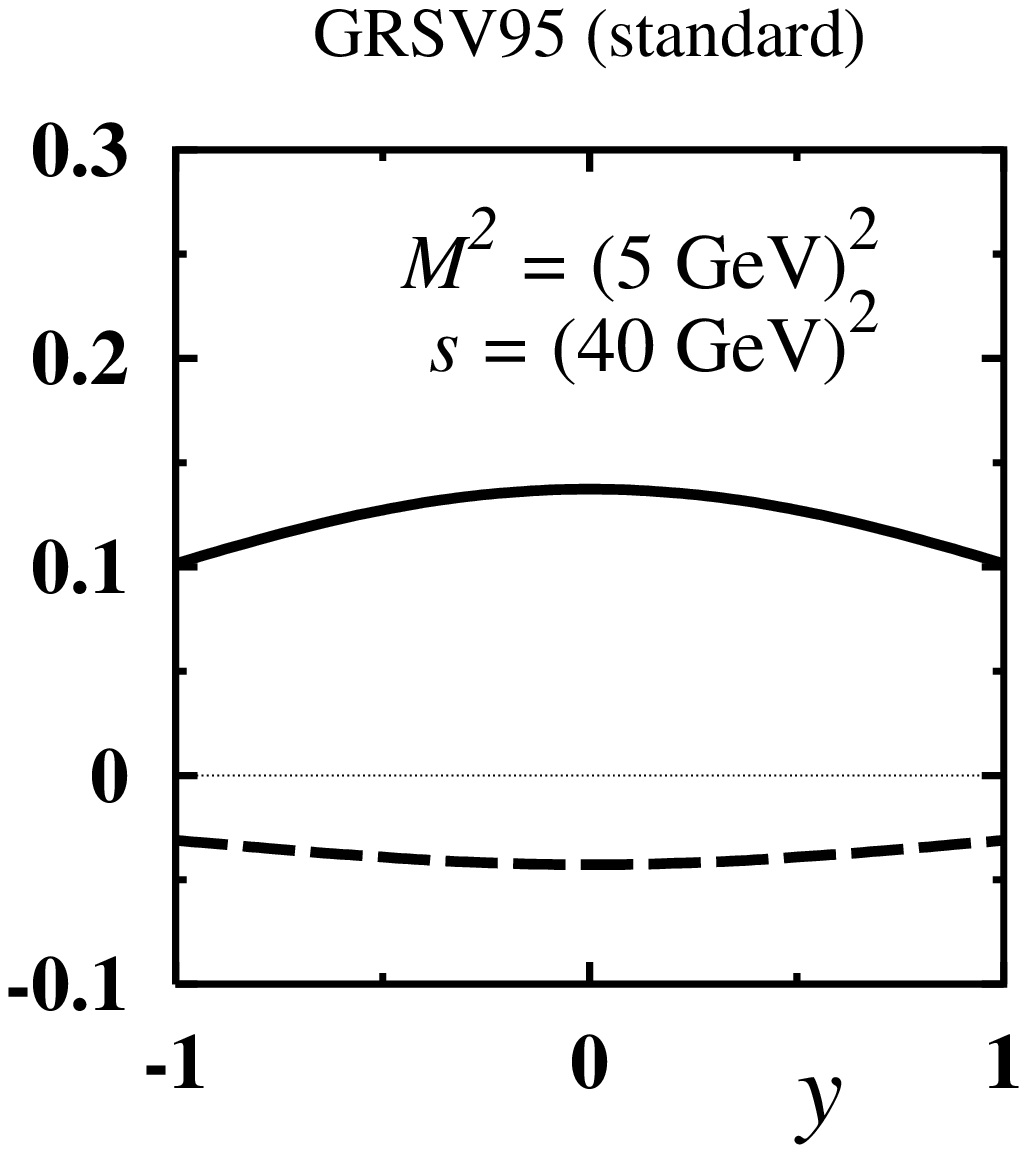}
\\
{\bf\Large (a)}
\end{tabular}
&
\begin{tabular}{c}
\includegraphics[width=7.2cm, height=7.2cm]{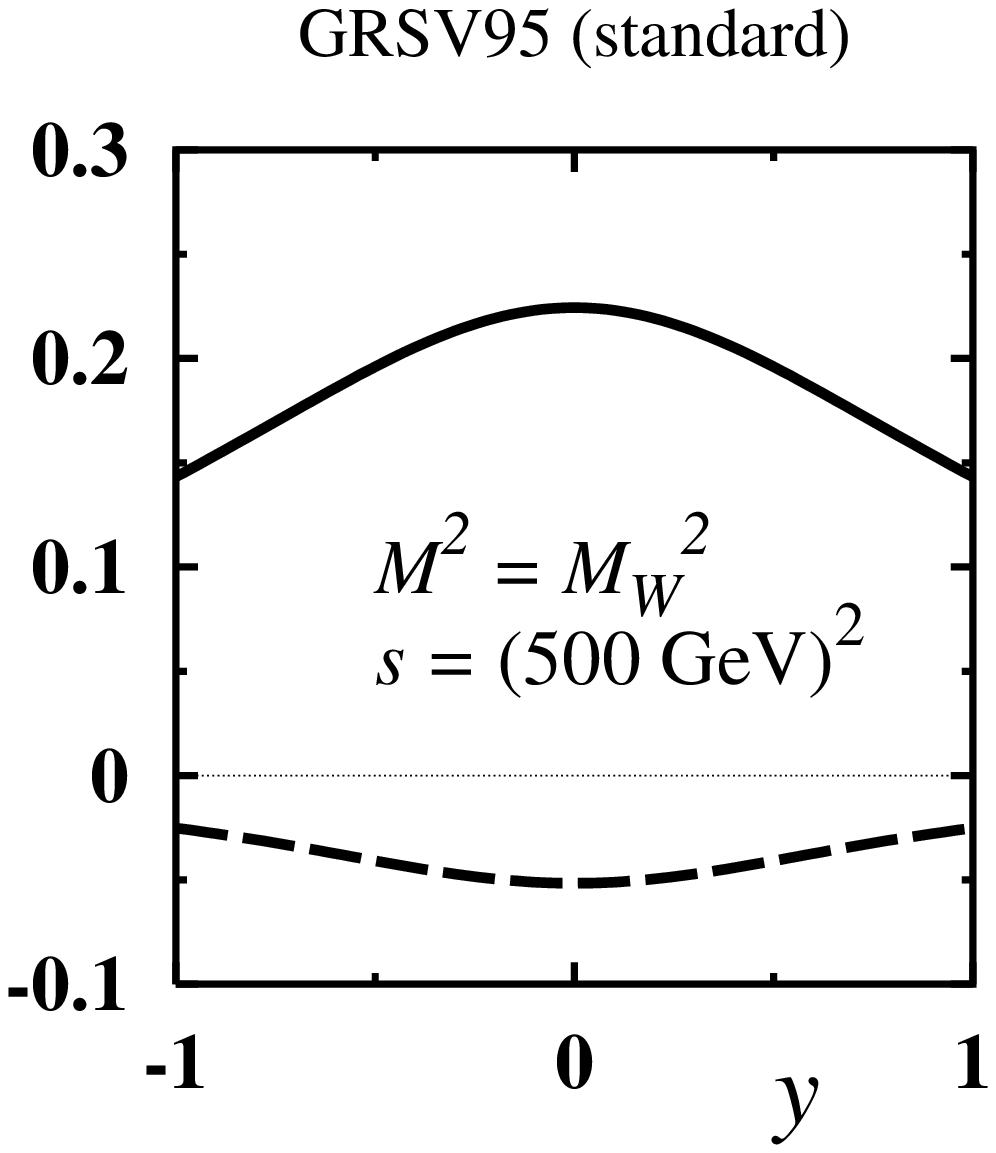}
\\
{\bf\Large (b)}
\end{tabular}
\\[2cm]
\begin{tabular}{c}
\includegraphics[width=7.2cm, height=7.2cm]{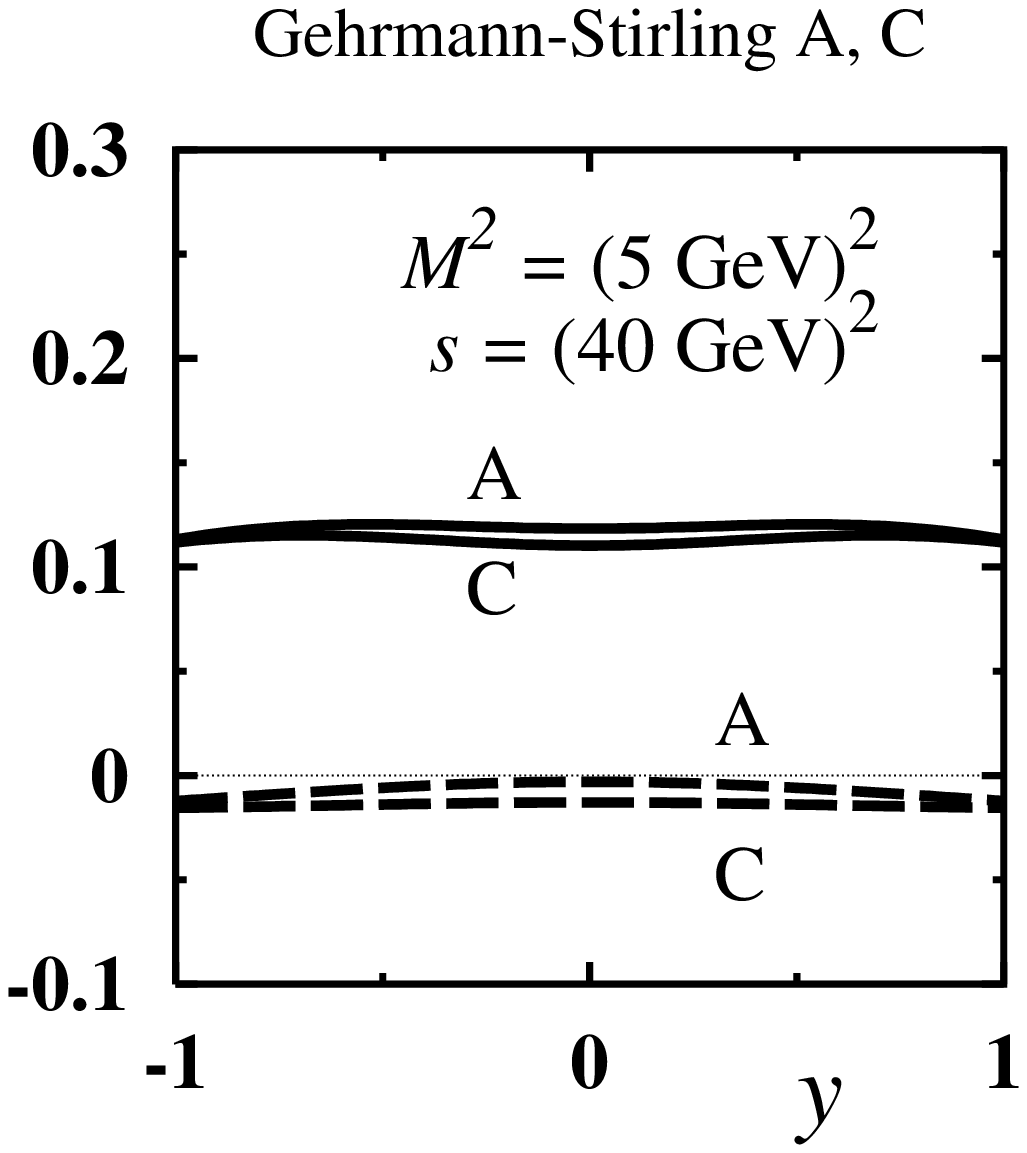}
\\
{\bf\Large (c)}
\end{tabular}
&
\begin{tabular}{c}
\includegraphics[width=7.2cm, height=7.2cm]{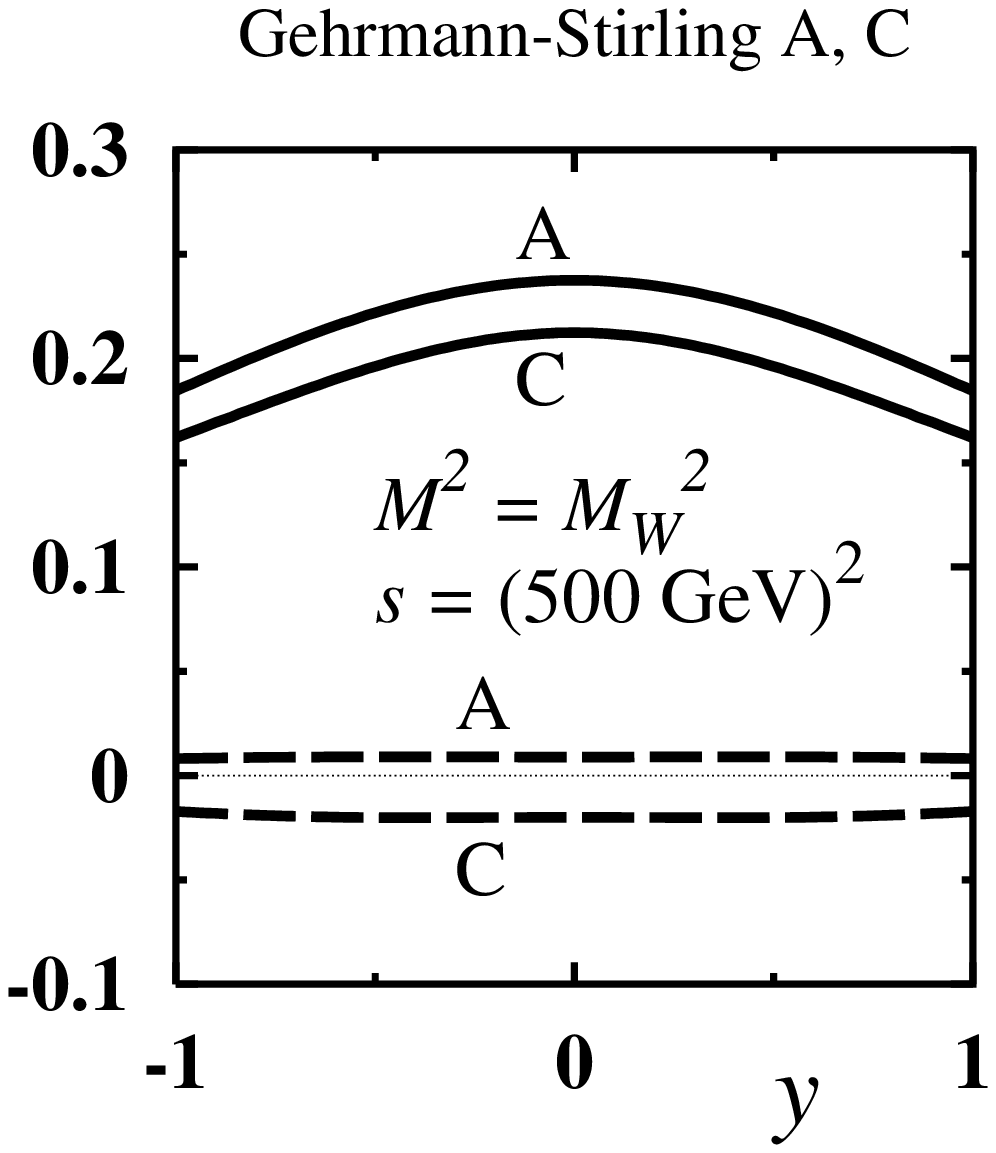}
\\
{\bf\Large (d)}
\end{tabular}
\end{tabular}
\caption[]{The longitudinal double spin asymmetry in DY pair
production through a virtual photon, $A^\gamma_{LL}$, in proton--proton
collisions, as a function of the
rapidity, $y$. Shown are the results for two different 
kinematical regions: $s = (40 \,{\rm GeV})^2, M^2 = (5 \,{\rm GeV})^2$ 
(HERA proton beam fixed--target experiment) and 
$s = (500 \,{\rm GeV})^2, M^2 = M_W^2 = (80.3 \,{\rm GeV})^2$ (RHIC).
\framebox{\rm (a), (b):}
\underline{Dashed lines:}
Results obtained for zero flavor asymmetry of the polarized
antiquark distributions, $\Delta_3 (x) = \Delta_8 (x) = 0$,
using the GRSV95 LO parametrizations {\rm \cite{GRSV96}}
for $\Delta_u (x), \Delta_d (x), \Delta_s (x)$ and $\Delta_0 (x)$.
\underline{Solid lines:} Results obtained including in addition
the antiquark flavor asymmetries, $\Delta_3 (x)$ and $\Delta_8 (x)$,
obtained in model calculations in the large--$N_c$ limit 
{\rm \cite{DPPPW96,Dressler99}}.
\framebox{\rm (c), (d):} same as
{\rm (a)} and {\rm (b)}, but using instead of GRSV95 the
Gehrmann--Stirling A and C parametrizations
{\rm \cite{GS96}}.}
\label{fig_A_LL}
\end{figure}

In Fig.\ref{fig_A_LL} (a) and (b) we compare the double spin asymmetries,
$A^\gamma_{LL}$, obtained with the ``hybrid'' distributions incorporating
the antiquark flavor asymmetries, $\Delta_3 (x)$ and $\Delta_8 (x)$, calculated
in the large--$N_c$ limit (solid lines), with what one obtains for
$\Delta_3 (x) = \Delta_8 (x) = 0$ (dashed lines). We show the results
in two different kinematical regions, (a): $s = (40\, {\rm GeV})^2$
and $M^2 = (5\, {\rm GeV})^2$, corresponding to a proposed
fixed target experiment using the HERA proton beam \cite{HERA95}, and
(b): $s = (500\, {\rm GeV})^2$ and $M^2 = M_W^2 = (80.3\, {\rm GeV})^2$,
which can be reached in the RHIC experiment. One sees that in both cases
the flavor asymmetry of the antiquark distribution
has a dramatic effect on the spin asymmetry, reversing even its sign
compared to the case with $\Delta_3 (x) = \Delta_8 (x) = 0$.
\par
The results for the double spin asymmetry, $A^\gamma_{LL}$,
depend in principle also on the assumptions made about the polarized
gluon distribution in the nucleon, which mixes with the singlet
quark distribution under evolution, and which is practically not
constrained by the present data. In order to estimate the sensitivity
of our results to the polarized gluon distribution
we have repeated the above comparison using instead of GRSV95 the
Gehrmann--Stirling LO ``A'' and ``C'' parametrizations for
$\Delta_u, \Delta_d, \Delta_s$ and $\Delta_0$, which provide
fits to the inclusive data with widely different assumptions 
about the shape of the input polarized gluon distributions \cite{GS96}. 
The resulting asymmetries $A^\gamma_{LL}$ obtained
without polarized flavor asymmetry, $\Delta_3 (x) = \Delta_8 (x) = 0$ 
(dashed lines), and including the large--$N_c$ model results 
for $\Delta_3 (x)$ and $\Delta_8 (x)$ (solid lines) 
are shown in Fig.\ref{fig_A_LL} (c) and (d).
One sees that the changes of $A^\gamma_{LL}$ due to the inclusion of the
flavor asymmetry (differences between corresponding solid and dashed curves)
are much larger than the differences due to changes of the input gluon
distribution (differences between the two dashed curves).
It is not an exaggeration to say that $A^\gamma_{LL}$ measures the
polarized flavor asymmetry of the antiquark distribution, and not the 
polarized gluon distribution. 
\par
Our comparison of asymmetries calculated with and without inclusion
of a polarized antiquark flavor asymmetry refers explicitly to the
leading--logarithmic (LO) approximation, since only at this level
the flavor asymmetries $\Delta_3 (x)$ and $\Delta_8 (x)$, evolve
separately and can be combined with parametrizations for 
$\Delta_u, \Delta_d , \Delta_s$ and $\Delta_0$ without affecting
the fits to inclusive data. It is expected that the spin 
asymmetry $A^\gamma_{LL}$ is less sensitive to NLO corrections
than the polarized and unpolarized DY cross sections individually,
since the $K$--factors partially cancel between numerator and denominator
in the ratio, Eq.(\ref{A_LL_parton}) \cite{Ratcliffe82}; however, 
this claim has been 
debated in Ref.\cite{Gehrmann97}. In any case, since the inclusion 
of the polarized antiquark flavor asymmetry has a very large effect 
on $A^\gamma_{LL}$ already at LO level, it is unlikely that higher--order 
corrections will reverse this situation. At least, the differences between
our LO results for $A^\gamma_{LL}$ obtained with and without flavor asymmetry
are much larger than those between the LO and NLO results in the case
of zero flavor asymmetry quoted in Ref.\cite{Gehrmann97}.
\par
\begin{figure}[t]
\begin{tabular}{rr}
\begin{tabular}{c}
\includegraphics[width=7.2cm, height=7.2cm]{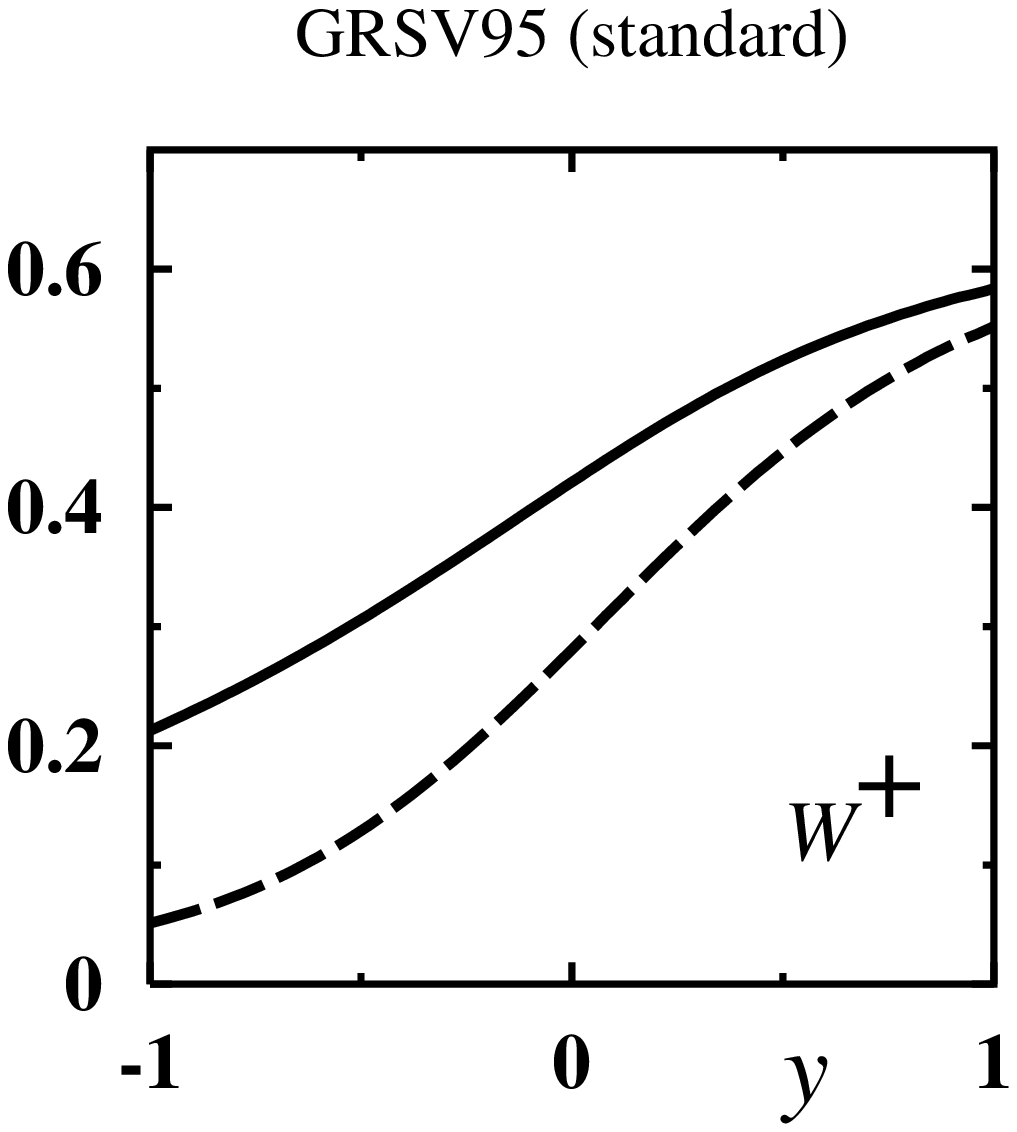}
\\
{\bf\Large (a)}
\end{tabular}
&
\begin{tabular}{c}
\includegraphics[width=7.2cm, height=7.2cm]{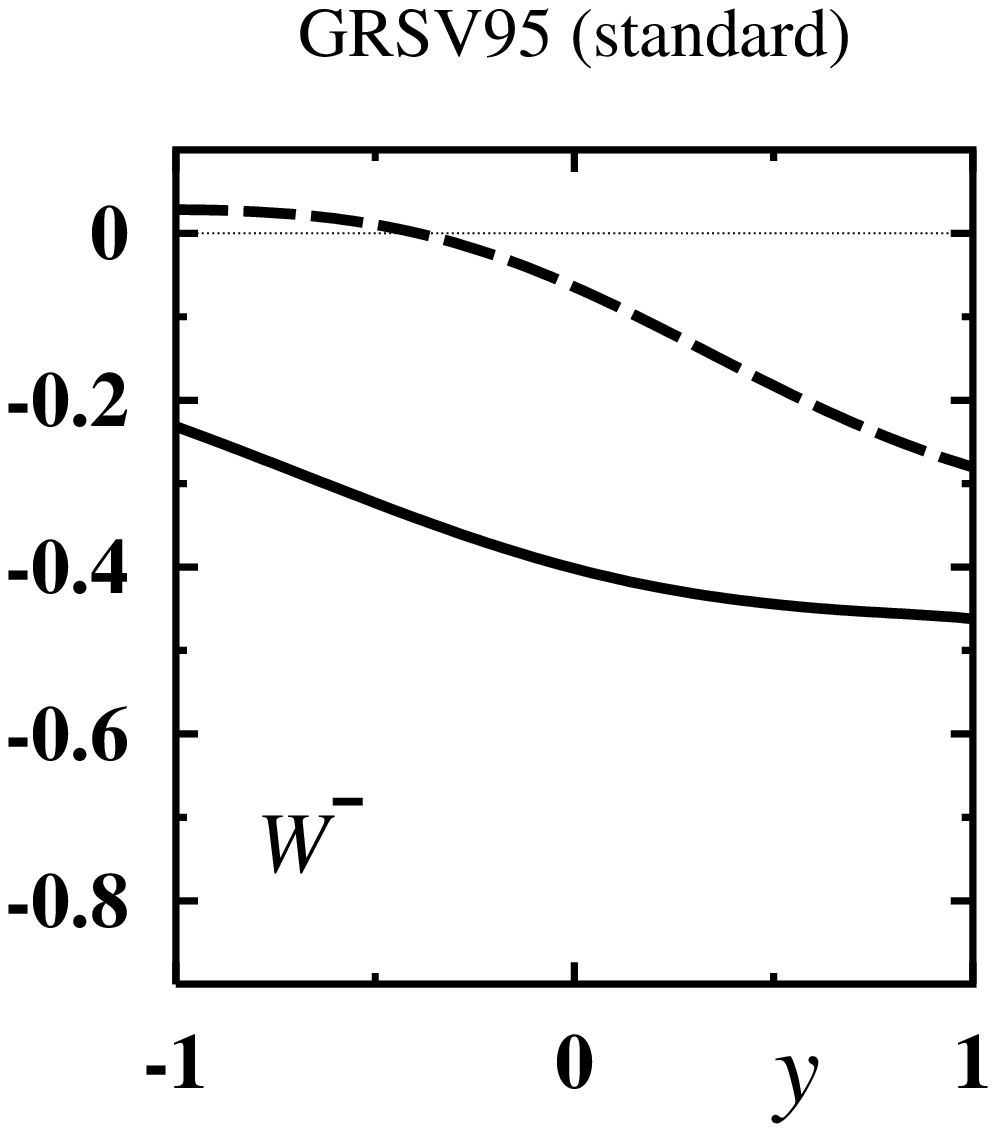}
\\
{\bf\Large (b)}
\end{tabular}
\\[2cm]
\begin{tabular}{c}
\includegraphics[width=7.2cm, height=7.2cm]{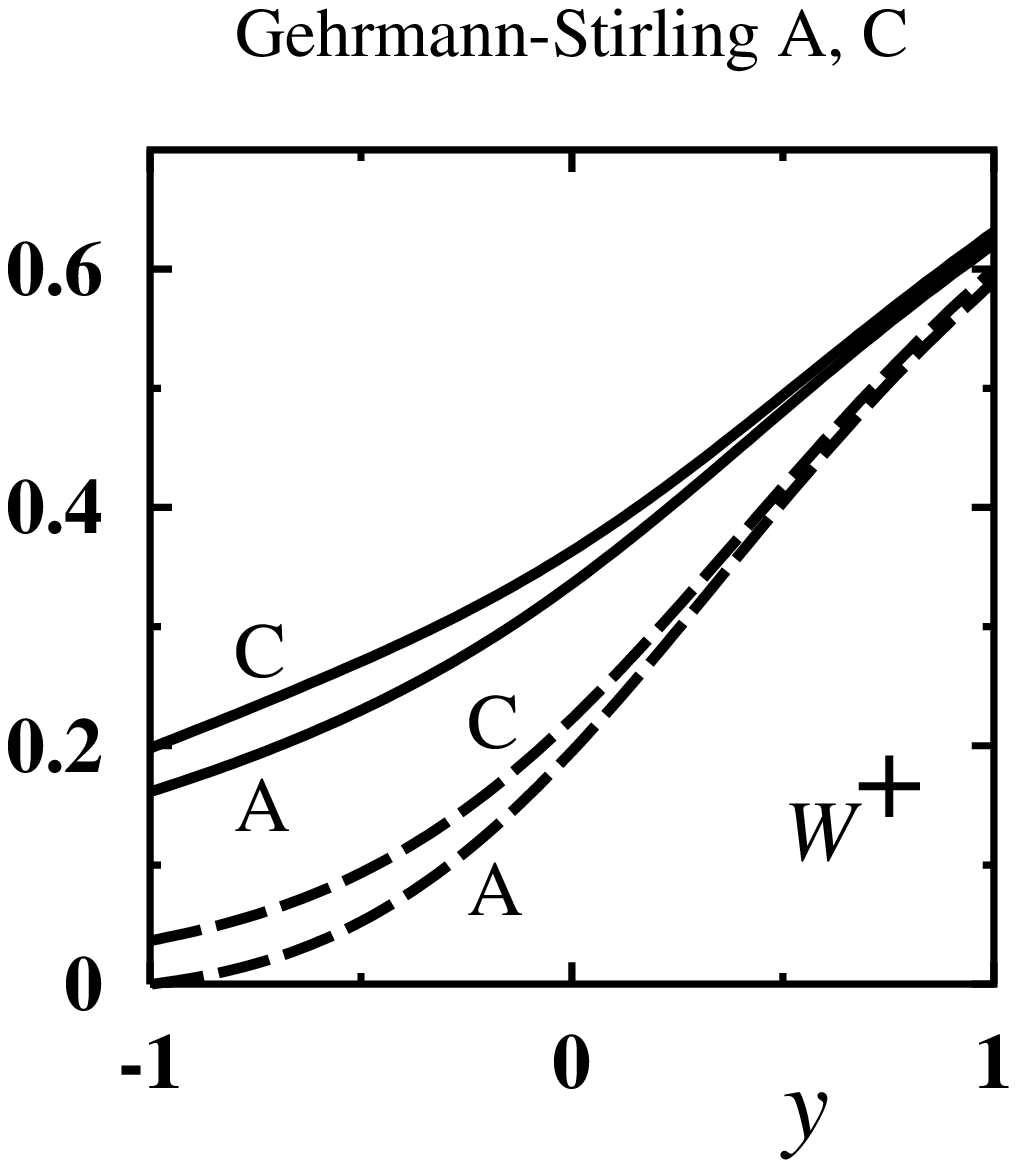}
\\
{\bf\Large (c)}
\end{tabular}
&
\begin{tabular}{c}
\includegraphics[width=7.2cm, height=7.2cm]{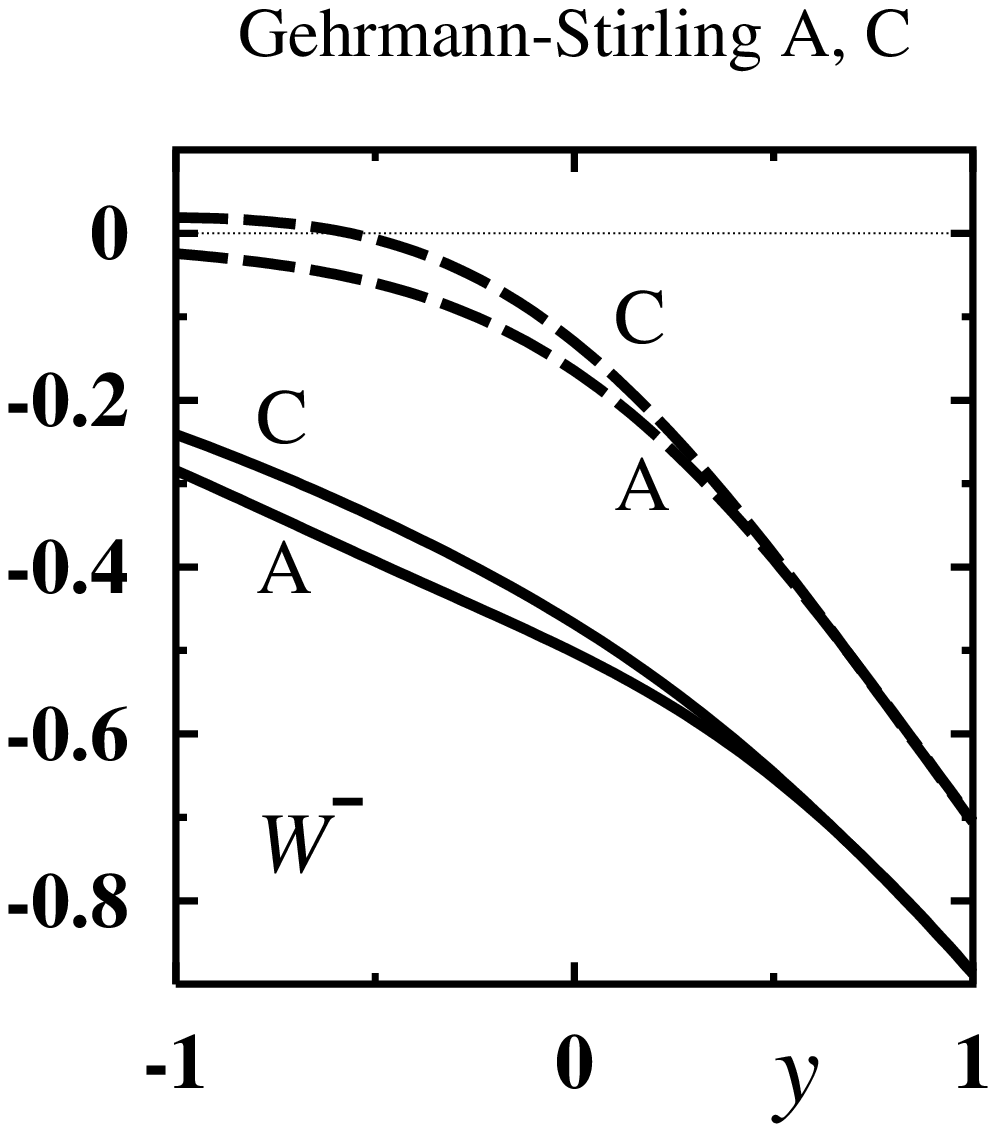}
\\
{\bf\Large (d)}
\end{tabular}
\end{tabular}
\caption[]{The longitudinal single spin asymmetry in lepton pair 
production through $W^+$ and $W^-$ bosons, $A^{W+}_{L}$ and $A^{W-}_{L}$,
in proton--proton collisions, as a function of the
rapidity, $y$, for $M^2 = M_W^2 = (80.3 \, {\rm GeV})^2$
and $s = (500\, {\rm GeV})^2$.
\framebox{\rm (a), (b):}
\underline{Dashed lines:}
Results obtained for zero flavor asymmetry of the polarized
antiquark distributions, $\Delta_3 (x) = \Delta_8 (x) = 0$,
using the GRSV95 LO parametrizations {\rm \cite{GRSV96}}
for $\Delta_u (x), \Delta_d (x), \Delta_s (x)$ and $\Delta_0 (x)$.
\underline{Solid lines:} Results obtained including in addition
the antiquark flavor asymmetries, $\Delta_3 (x)$ and $\Delta_8 (x)$,
obtained in model calculations in the large--$N_c$ limit 
{\rm \cite{DPPPW96}}.
\framebox{\rm (c), (d):} same as
{\rm (a)} and {\rm (b)}, but using instead of GRSV95 the
Gehrmann--Stirling A and C parametrizations
{\rm \cite{GS96}}.}
\label{fig_A_L}
\end{figure}

The single spin asymmetries in lepton pair production through
$W^\pm$, $A^{W\pm}_L$, for proton--proton scattering are
shown in Fig.\ref{fig_A_L}, for
$s = (500\, {\rm GeV})^2$ and $M^2 = M_W^2 = (80.3\, {\rm GeV})^2$,
which can be reached at RHIC.
Figs.\ref{fig_A_L} (a) and (b) show the results obtained using the
GRSV95 parametrization without antiquark flavor asymmetry
(dashed lines), and including the contributions from $\Delta_3 (x)$
and $\Delta_8 (x)$ obtained in the large--$N_c$ model estimate 
\cite{DPPPW96,Dressler99} (solid lines). 
One sees that also in this case the inclusion of the antiquark flavor
asymmetry has a qualitative effect on the spin asymmetry.
Again, in the case of the
Gehrmann--Stirling parametrizations, Fig.\ref{fig_A_L} (c) and (d),
the differences due to changes in the gluon distribution are negligible
compared to the effect of the flavor asymmetry of the antiquark distribution.
\par
To summarize, we have shown that the large flavor
asymmetries of the polarized antiquark distributions predicted by
model calculations in the large--$N_c$ limit (chiral quark--soliton model), 
have a pronounced effect on the spin asymmetries in Drell--Yan pair 
production through photons or $W^\pm$ bosons at HERA or RHIC energies.
In particular, the effect of the antiquark flavor asymmetry on
the spin asymmetries is much larger than their uncertainties due
to the lack of knowledge of the degree of gluon polarization in the
nucleon. The expected accuracy of the RHIC measurements \cite{Heppelmann}
will certainly be sufficient to observe an effect of the magnitude
predicted.
\\[.5cm]
We are grateful to S.\ Heppelmann and P.V.\ Pobylitsa for useful 
discussions. This investigation was supported in part by the Deutsche 
Forschungsgemeinschaft (DFG), by a joint grant of the 
DFG and the Russian Foundation for Basic Research, by
the German Ministry of Education and Research (BMBF),
and by COSY, J\"ulich. The work of M. Strikman was supported 
in part by a DOE grant, and by the Alexander--von--Humboldt 
Foundation.
%

%
%
\newpage
%
%
\newpage
%
%
\newpage
\end{document}